\documentclass{sigchi}

\toappear{Permission to make digital or hard copies of all or part of this work for personal or classroom use is granted without fee provided that copies are not made or distributed for profit or commercial advantage and that copies bear this notice and the full citation on the first page.} 

\usepackage{balance}  
\usepackage{graphics} 
\usepackage[T1]{fontenc}
\usepackage{txfonts}
\usepackage{mathptmx}
\usepackage[pdftex]{hyperref}
\usepackage{color}
\usepackage{booktabs}
\usepackage{textcomp}
\usepackage{microtype} 
\usepackage{ccicons}  
\usepackage[utf8]{inputenc} 

\usepackage{todonotes}

\def\plaintitle{Writers Gonna Wait: The Effectiveness of Notifications to Initiate Aversive Action in Writing Procrastination}

\def\emptyauthor{}
\def\plainkeywords{Procrastination, Behavior Change, Writing}

\makeatletter
\def\url@leostyle{%
  \@ifundefined{selectfont}{
    \def\UrlFont{\sf}
  }{
    \def\UrlFont{\small\bf\ttfamily}
  }}
\makeatother
\def\pprw{8.5in}
\def\pprh{11in}

\definecolor{linkColor}{RGB}{6,125,233}
\hypersetup{%
  pdftitle={\plaintitle},
  pdfauthor={\emptyauthor},
  bookmarksnumbered,
  pdfstartview={FitH},
  colorlinks,
  citecolor=black,
  filecolor=black,
  linkcolor=black,
  urlcolor=linkColor,
  breaklinks=true,
}

\begin{document}

\title{\plaintitle}

\numberofauthors{3}
\author{%
  \alignauthor{Chatchai Wangwiwattana\\
    \affaddr{Southern Methodist University, Department of Computer Science}\\
    \email{cwangwiwatta@smu.edu}}\\
  \alignauthor{Sunjoli Aggarwal\\
    \affaddr{Southern Methodist University, Department of Computer Science}\\
    \email{sunjolia@smu.edu}}\\
  \alignauthor{Eric C. Larson\\
    \affaddr{Southern Methodist University, Department of Computer Science}\\
    \email{eclarson@smu.edu}}\\
}

\maketitle

\begin{abstract}
This paper evaluates the use of notifications to reduce aversive-task-procrastination by helping initiate action. Specifically, we focus on aversion to graded writing tasks. We evaluate software designs commonly used by behavior change applications, such as goal setting and action support systems. We conduct a two-phase control trial experiment with 21 college students tasked to write two 3000-word writing assignments (14 students fully completed the experiment). Participants use a customized text editor designed to continuously collect writing behavior. The results from the study reveal that notifications have minimal effect in encouraging users to get started. They can also increase negative effects on participants. Other techniques, such as eliminating distraction and showing simple writing statistics, yield higher satisfaction among participants as they complete the writing task. Furthermore, the incorporation of text mining decreases aversion to the task and helps participants overcome writer's block. Finally, we discuss lessons learned from our evaluation that help quantify the difficulty of behavior change for writing procrastination, with emphasis on goals for the HCI community.

\end{abstract}
\category{H.5.m.}{Information Interfaces and Presentation
  (e.g. HCI)}{Miscellaneous} 


\section{Introduction}
For decades, the HCI community has researched persuasive design in behavior change in applications ranging from health improvement \cite{klasnja_how_2011,larson_spirosmart:_2012,madan_social_2010}, to well-being \cite{lane_bewell:_2014,rabbi_passive_2011}, to sustainability \cite{larson_disaggregated_2012,lehrer_evaluating_2011}. These researchers seek to bridge the gaps between practical HCI design and behavioral psychology---or, alternatively, behavioral economics or neuroscience; nonetheless, this gap has proven difficult to trellis \cite{Zimmerman2007ResearchHCI}. Because of the generalized nature of behavioral theory, there are many possible ways to apply the same knowledge in practical applications \cite{Zimmerman2007ResearchHCI}. That is, a technique that works well within one context may not work in another. 

One widely used mechanism in behavior change is the notification. The notification is used to draw users'  attention to a task or part of a task with the hope that action will be initiated. In theory, a psychological trigger, whether internal or external, is the first step of any behavior \cite{Wendel2013DesigningEconomics}. Notifications, therefore, can be categorized as effective external cues to initiate behavior \cite{Eyal2014Hooked:Products}. However, many studies investigate the use of notifications for tasks that users already have an internal motivation to complete; tasks for which internal motivation is absent have not been studied as thoroughly. Such tasks are also known as ``aversive tasks.'' High intention to perform aversive tasks does not guarantee that the behavior will occur \cite{ajzen_attitude-behavior_1977}. Thus, it is an open question to what degree notifications are effective psychological triggers for completing a task. In this paper, we evaluate the usage of notifications for changing behavior towards aversive tasks, in which there is no internal motivation present. More specifically, we focus on reducing writing procrastination among college students.

We outline our contributions as follows: (i) We investigate the role of  human computer interaction in formal psychological theories of  procrastination behavior. (ii) We evaluate the efficacy of various notification styles in reducing procrastination of tedious writing tasks. (iii) We evaluate popular text editor designs and their perceived effects on procrastination behavior. (iv) We summarize lessons learned from working on procrastination research from an HCI perspective.

Writing is considered an unpleasant activity by many individuals because it requires tremendous effort, is susceptible to judgment, and has delayed reward; nonetheless it is vital to succeeding in many professions and, therefore, cannot be treated as an optional lifestyle choice. The conflict of \textit{have to} and \textit{want to}, which psychologists call cognitive dissonance, elicits negative effects such as guilt and  distress \cite{Ferrari1994DysfunctionalBehaviors}. These negative effects enter into a positive feedback loop---the more a student procrastinates, the worse he/she feels, and these negative feelings block him/her from having an environment conducive to writing, leading to further procrastination \cite{ferrari_procrastination_2013}. Due to its cyclic nature,  writing procrastination is an extremely challenging  application of aversive behavior change research, but as stated, trying to reduce this phenomenon is also extremely relevant.   


In this paper, we evaluate the effectiveness of notifications and other persuasive HCI design elements on reducing procrastination and initiating action for writing tasks. We developed a research instrument for collecting data and facilitating the experiment (\textit{i.e.}, a custom text editor that tracks interaction and can distribute self-evaluation surveys). The custom text editor has a number of features, including a notification system, a goal progress bar, distraction-free mode, and a writing assistance system. All of these features are  designed based on psychological intervention theory. We conducted an experiment with 21 college students who used the editor to complete graded writing assignments. The evaluation lasted for eighteen days. Students worked on two separate writing assignments: the first assignment lasted for nine days without notifications and was used as a baseline evaluation. In a second nine day experiment, we divided the same users into two groups based on the type of notifications they received (\textit{normal} or \textit{actioned} notifications). We recorded users' progress using word count, writing time, and start time, and recorded users' interaction with notifications for both assignments. We also collected self-reported survey data on procrastination with PASS (Procrastination Assessment Scale), and user satisfaction with SUS (System Usability Scale). In the follow-up study, 14 participants used the editor regularly enough for analysis. The results showed minimal change in writing procrastination behavior regardless of the type or number of notifications received. Surprisingly, participants who received more notifications reported \textit{lower} satisfaction than others who received fewer notifications. In addition, we found our custom writing assistance system, which employs text mining and concept models to help mitigate writer's block, reduced aversion towards the task and provided a more positive writing experience. Other persuasive techniques, such as a simple goal progress bar and distraction-free mode, also showed positive outcomes. 

\section{Understanding Procrastination}
In order to bring the HCI community together around behavior change for procrastination, we  first describe the psychology community perspectives (which are competing in some instances). This also helps to properly ground our HCI methodologies in the context of current psychological research trends. 

A valid question to ask before going further is: \textit{Should we be trying to reduce procrastination behavior?} In general, social psychologists argue that procrastination is an irrational and maladaptive behavior \cite{ferrari_still_2010}. There is considerable evidence that procrastination not only harms productivity, but also increases stress and contributes to a poorer quality of life. Studies show that procrastination can affect people's health, ranging from headaches, body aches, and colds to tooth decay, stress, and strokes \cite{sirois_procrastination_2004,sirois_ill_2007,sirois_ill_2003}. Moreover, in academic literature, studies have shown that academic procrastinators are likely to engage in cheating and plagiarism, as well as have problems with self-esteem and self-confidence \cite{ferrari_still_2010}. Procrastinators also have a high degree of self-handicapping behaviors, such as indecision, rebelliousness,  societal demands for perfection, and a low degree of optimism, self-esteem, life satisfaction, and self-confidence \cite{ferrari_social_2004}. In contrast to these negative effects, some research suggests that procrastination exists for good reasons. From an evolutionary perspective, procrastination yields benefit because the long term goals that a non-procrastinator would focus on may actually distract from short term survival goals that a procrastinator would gravitate towards \cite{Gustavson2014GeneticProcrastination}. Norman argues that procrastination provides the maximum time to think, plan, and determine alternatives, giving more flexibility for change in the future and allowing requirement to evolve \cite{Norman2014WhyGood}. In this paper, we do not argue whether procrastination is or is not beneficial. Instead, we argue that individuals who wish to change their behavior should have a right to do so. Our research, thus, provides valuable knowledge  for individuals who wish to reduce their procrastination behavior using technology aides.

Procrastination can be viewed from various perspectives. Cognitive Science views procrastination as a subtle executive dysfunction \cite{rabin_academic_2011}. Executive functions rely on a number of interconnected cortical and sub-cortical brain regions. Together, these areas are responsible for the self-regulation of cognition and for all cognitive processes that enable planning for complex actions \cite{rabin_academic_2011}. In contrast to Cognitive Scientists, Evolutionary Psychologists view procrastination as a result of human evolution. They believe that focusing on short term survival results in a greater chance of passing on a gene, whereas long term planning is merely a distraction. According to them, procrastination is an evolutionary by-product of impulsiveness \cite{Gustavson2014GeneticProcrastination}. Social Psychologists have a similar view to Cognitive Scientists. Piers Steel defined procrastination as ``to voluntarily delay an intended course of action despite expecting to be worse off for the delay'' \cite{steel_nature_2007}. To emphasize its negative nature, Kyle further defines procrastination as ``the voluntary, irrational postponement of an intended course of action despite the knowledge that this delay will come at a cost to or have negative affects on the individual'' \cite{kyle_search_2009}. In short, many psychologists agree that procrastination is a form of self-regulation failure. It is important to note that not all delays are categorized as procrastination. For example, a planned delay or a delay from external factors is \textit{not} procrastination; procrastination must include an \textit{irrational} delay. With this definition, Norman's argument---to maximize time to think, plan, and determine alternatives, give more flexibility in future change, and allow requirements to change\cite{Norman2014WhyGood})---would not be considered procrastination; it is time management. 

\subsection{Characteristics of Procrastinators}
Individuals who procrastinate often repeat procrastination behavior; therefore, the psychology community categorizes people as procrastinators and non-procrastinators. Procrastinators have unique characteristics compared to non-procrastinators. For example, many procrastinators have a misbelief that pressure motivates them to do their best work \cite{kyle_search_2009}; furthermore, procrastinators have high sensitivity to immediate gratification and have trouble focusing on tasks \cite{ferrari_procrastination_2000,ferrari_procrastination_2001,ferrari_regulating_2007}. Procrastinators also typically have low self-control and low self-reinforcement, meaning they are unable to reward themselves for success \cite{ferrari_methods_1995}. They also have decreased ability to regulate their performance or speed under restricted time frames \cite{ferrari_procrastination_2001} and reflect on the future negatively. In the specific case of students, distractions come easy \cite{ferrari_psychometric_1992, ferrari_procrastination_2000, ferrari_frequent_2007} and the ability to estimate the amount of time necessary to finish a task is lacking \cite{ferrari_procrastination_2013}.

\subsection{Causes of Procrastination}
Causes of procrastination are complex. Procrastination can stem from the task itself or from individual differences in terms of personality and genetic uniqueness. Task Aversion, \textit{dysphoric affect} \cite{Milgram1993CorrelatesProcrastination}, or task appeal \cite{Harris1983TaskResearch} refer to an action that one finds unpleasant. By this definition, the more aversive the task, the more likely one is to avoid it. Timing, rewards, and punishments also influence the path towards procrastination. This is known in the behavioral economics community as \textit{intertemporal choice} or \textit{discounted utility} \cite{Loewenstein1992ChoiceTime}. Marketing researchers view procrastination as deciding to perform a certain task in a certain amount of time based on the perfect compromise between cost and value. To a procrastinator, present cost is usually perceived as higher than future costs, while value remains constant. This leads one to act closer to the deadline, even though the action may be an enjoyable activity \cite{SHU2013ProcrastinationExperiences}.

Social psychologists argue that procrastination is stimulated by negative causes. For example, fear of failure can contribute to procrastination \cite{schouwenburg_procrastinators_1992}. Ferrari, based on 20 years of research, proposed three models of procrastination: Arousal, Avoidant, and Decisional. \textit{Avoidant procrastinators} have the tendency to avoid certain outcomes such as fear of failure, success, social isolation, or feeling like an impostor. \textit{Arousal procrastinators} rely on pressure in order to work. \textit{Indecisive procrastinators} intentionally decide not to act \cite{ferrari_aarp_2011, ferrari_experimental_1997}. Indecisive procrastination is related to a lack of competence or time urgency. It is not related to laziness, but is rather more about not understanding the trade-off between speed and accuracy \cite{ferrari_experimental_1997}. Ferrari also argues that \textit{learned helplessness}, or the situation of experiencing a series of uncontrollable and unpredictable unpleasant events \cite{seligman_prevention_1999}, contributes to procrastination. For example, some procrastinators use procrastination as a self-handicapping strategy. When procrastinators perceive low-competence, they blame external factors (such as not having enough time) in order to protect their self-esteem and themselves from social judgment for poor performance. 

 
The behavior-intention gap, addressed in \textit{Theory of Reasoned Action}, is where there is a gap between behavioral intention and behavior. Intention is a strong predictor of behavior; however, that behavior is not guaranteed even when the person believes that the act is worthwhile \cite{ajzen_theories_1991, ajzen_attitude-behavior_1977}. Enjoyable activities that can be done right away have more $utility$ than non-urgent or undesirable tasks. This result was supported by Haghbin \textit{et al.} \cite{haghbin_complexity_2012}. 

\subsection{Possible Psychological Interventions} \label{sec-interventions}
Examples of psychological interventions to overcome procrastination are plentiful. A popular intervention is isolating a task and breaking it into small and attainable steps \cite{Haycock1998ProcrastinationAnxiety}. The effect can be enhanced by incorporating goal setting theory, entailing the creation of small sub goals and enforcing regular deadlines. This helps users regain self-efficacy and narrow the intention-action gap \cite{steel_nature_2007}. In addition, Ainslie and Haslam suggest training procrastinators to separate negative effects from taking action \cite{Loewenstein1992ChoiceTime}. This can be seen as a combination of \textit{willpower} and \textit{mindfulness training}. Schouwenburg suggests using \textit{commitment devices} to limit and eliminate short-term temptations altogether such as turning on \textit{do not disturb mode} or disconnecting the Internet \cite{Schouwenburg1995AcademicResearch}. Instead of trying to rely on willpower or commitment devices, \textit{Implementation Intention} exploits psychological triggers and the power of habit. It consists of forming an intention with specific action plans. That intention acts as a cue for triggering followed behaviors. For example, \textit{If I go to a restroom, I will also get a cup of water.} A study shows hat this makes individuals nearly eight times more likely to follow through with a task \cite{owens_overcoming_2008}. \textit{Time Traveling}, in contrast, focuses on mood regulation. This approach encourages individuals to flip negative and positive reflections toward tasks by asking them to think about how they will feel after the task is completed \cite{pychyl_solving_2013}.  Recent work in psychology explores the idea of having individuals make a relationships with their future-selves to help procrastinators make more rational decisions \cite{gilbert_psychology_2014,goldstein_battle_2001}. The idea of the future-self focuses on how current decisions will affect one's self in the future. For an academic setting, Ferrari suggests (1) finding the part of the paper with the most individual interest, (2) creating an outline, and (3) writing in small sessions \cite{ferrari_procrastination_2013}. Pychyl believes that just getting started is the most effective way to decrease procrastination \cite{pychyl_solving_2013}, arguing that splitting the task into small, manageable sub-goals, it becomes easier to start.

Although these techniques have reported success, one limitation is the training and active commitment required to implement them into one's life. Like other behavioral interventions, ``only knowing how'' does not cause changes in behavior \cite{schultz_knowledge_2002}.  Even through one may implemented it, it is still unclear how sustainable the initial commitment will last. We argue that exploring how technology can augment psychological interventions is a vital role for the HCI community. In the study presented here, we build the interventions explicitly into our text editor system, hypothesizing that the system might guide productive behavior if it is a part of the holistic intervention process.

\section{Is Technology a New Thief of Time?}
Joseph Ferrari condemns modern technology in his book, ``Technology the New Thief of Time'' \cite{ferrari_procrastination_2013}. There is evidence to support his claim. In 2008, office workers spent 28\% of their time managing technology interruption and 46 percent of those interruptions — nearly half — were not necessary and not urgent \cite{ferrari_aarp_2011}. This report does not include the increasing number of notifications from applications that demand attention. Cyber-psychologists have coined this ``e-procrastination'' and associate this with ``attitudes like a sense of low control over one's time and a desire to take risks by visiting Web sites that are forbidden'' \cite{thatcher_online_2008}. Although this evidence positions modern technology in a bad light, it does not represent the entire story. Technology can also be a tool, depending on how we use it. Some practitioners recognize the problem of procrastination and offer various solutions. ``Stop Procrastination,'' for example, is an application that blocks distracting sites and emails \cite{_stop_2016}. It is designed to eliminate interruptions, which has some positive result \cite{patterson_can_2014}.  ``Avoiding Procrastination 101'' teaches users various techniques about procrastination \cite{_101_2012}. ``Write or Die 2'' allows users to choose either negative or positive consequences when they lose focus \cite{_write_2016}. However, there has been little evidence to support whether or not these techniques work. In this paper, we evaluate some popular technological techniques of behavior change: triggers, eliminating distraction, splitting to smaller sub-tasks, goal setting, and machine augmented intelligence.

\section{From Theories to UI Elements}
\subsection{Notifications for Getting Started}
Push notifications are common in behavior change applications. Studies show some types of notifications are capable of creating behavior and users appreciate having notifications as reminders \cite{bentley_power_2013}; nonetheless, not all notifications are equally important, and users are more likely to react to important notifications \cite{sahami_shirazi_large-scale_2014}. Eyal, the author of \textit{Hook}, argues that a good trigger should be well timed, actionable, and spark interest \cite{bentley_power_2013,Eyal2014Hooked:Products}. Interestingly, these qualities are found challenging in aversive tasks such academic writing. Aversive tasks are unlikely to spark authors' interest. They have the perception of requiring huge time and effort, and \textit{in-a-mood} writing time is unpredictable. We could say the notifications from aversive tasks are \textit{aversive notifications}---no one wants to get one, because it is a reminder of an aversive task. This leads us to ask, \textit{can notifications trigger actions in aversive tasks?}.

To answer this question, we implemented standard clickable notifications in our customized editor. We utilized psychological intervention and persuasive techniques to guide the content of the notifications. Moreover, we compare it with a new type of  notification we call an \textit{actioned notification}: focusing on eliciting an immediate action. 

\subsubsection{Standard Clickable Notifications}
\begin{figure}
\centering
\includegraphics[width=0.3\textwidth]{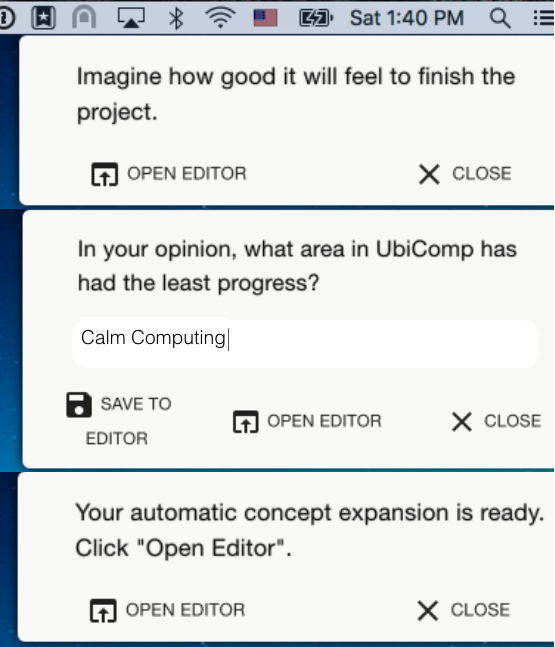}
\caption{\label{fig-popup}An example push notification (top), action notification with text prompt (middle), and concept expansion action notification (bottom).}
\end{figure}

We implement a standard clickable notification in our text editor that can contain various messages. Users can click on the notification to open the editor (see Figure \ref{fig-popup}). This notification is commonly used in behavior-change applications.  We use it to remind, inform, and motivate users. We group the messages into five categories based upon the intent of the notification:

\begin{enumerate}
\item \textbf{Standard Reminder}(e.g.,``It is your writing time.''): This type of notification only acts as a reminder. This concept is used in many task management applications such as to-dos and calendars. We require users to set their daily writing times, and they receive reminder notifications when the time is reached. A study shows that students who set their own deadline for their writing assignments perform better than those who do not \cite{Ariely2002ProcrastinationPrecommitment}. The strength of this notification type is that it is intuitive---most users are already familiar with this style of notification. 
\item \textbf{Encouraging Reminder} (e.g., ``Great Job! You have written 2000 words''): This type of notification is employed by many fitness and online learning applications. It attempts to increase user's motivation through positive reinforcement of previous activity.
\item \textbf{Inviting Action} (e.g.,``Let's write for 2 minutes!''): Unlike the encouraging reminder notification, \textit{Inviting Action} notifications attempt to create the perception that a task requires small effort.  Wendel defines \textit{Minimum Viable Action} to refer to the smallest action of the behavior. If the action is small enough, users are more likely to enact the behavior \cite{Wendel2013DesigningEconomics}
\item  \textbf{Tips and Tricks} (e.g.,``Writing Tip: An idea is nothing more nor less than a new combination of old elements. -- The Pareto Principle''): This type of notification is based on a \textit{Knowledge Deficit Model} \cite{schultz_knowledge_2002}. This model suggests that not knowing how to perform the behavior can block the behavior from happening, even though users have the right attitude. These notifications are meant to invoke thought about writing behavior via trying new suggested tips.
\item \textbf{Mood Regulation}  (e.g., ``Imagine how good it will feel to finish the project.''): This notification is based on the psychological intervention \textit{Time Traveling}. It is a mood regulation technique that attempts to convert negative reflections on doing the task to positive feelings about task completion.
\end{enumerate}

\subsubsection{Actioned Notifications}
Unlike standard clickable notifications, \textit{actioned notifications} encourage performing an action instead of attempting to increase motivation. We make the system split a task into small manageable chunks, and then present this chunk as a question-and-answer system. We hypothesized that this would reduce user's effort and increase the likelihood of the action occurring \cite{KingsleyZipf2016HumanEcology}. Actioned notifications contain a question and a text input box (see Figure \ref{fig-popup}). Users can answer the question right away in the text box. Furthermore, we hypothesized that the notification would help users focus on the \textit{action} rather than on the \textit{feelings} surrounding the action, potentially reducing aversion to the notification.

\begin{figure*}[h]
\centering
\includegraphics[width=\textwidth]{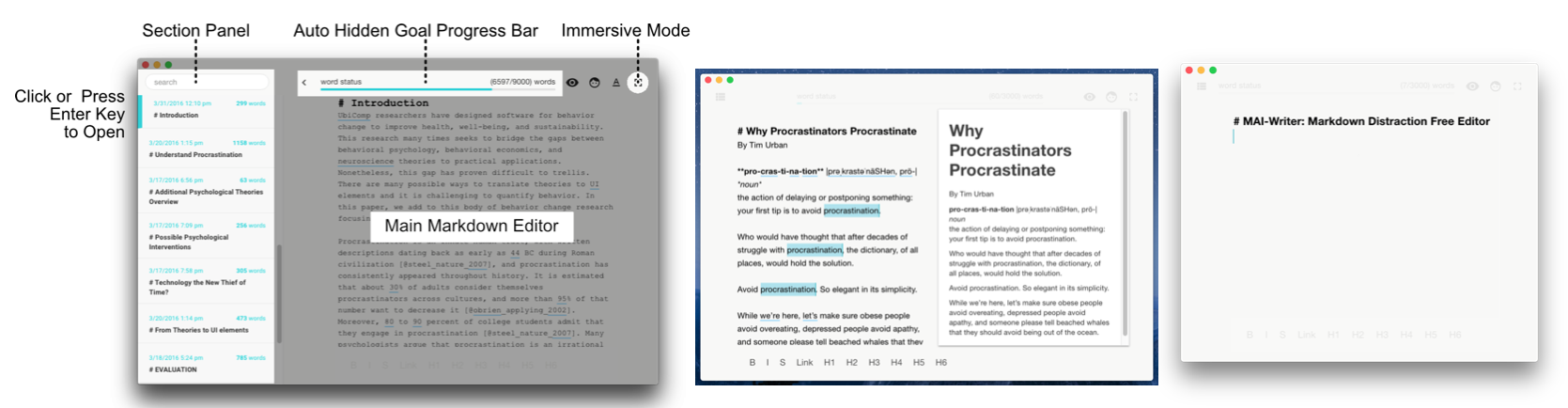}
\caption{\label{fig-maiUI}The main user interface for our custom text editor (left), markdown display (center), and distraction free mode (right).}
\end{figure*}

\subsection{Persuasive Elements for Retaining Behavior }
In addition to notifications, we designed our text editor with elements we hypothesized would be beneficial to reducing procrastination such as an immersive mode, a goal progress bar, and a writing assistance system.

\subsubsection{Immersive Mode}
This feature aims to curtail impulsiveness in procrastinators (see Figure \ref{fig-maiUI}). Studies show that procrastinators are sensitive to distraction \cite{ferrari_psychometric_1992, ferrari_procrastination_2000, ferrari_frequent_2007}. Immersive mode is a stage in which all external user interfaces are hidden and the editor expands itself full screen on top of other applications. Many text editors offer immersive mode. To further encourage users to focus on writing and not editing, the customized editor supports Markdown Language. Markdown allows the screen to be free from tools and buttons, making the interface simpler. All participants are computer science students, so the markdown format is familiar to them. Note that using markdown or immersive mode is optional.

\subsubsection{Goal Setting Theory: Goal Progress Bar}
Latham argues in his book, ``Goal setting theory is among the most valid and useful theories of motivation of organizational behavior'' \cite{LockeAPerformance}. Nonetheless, it is not a perfect solution. Goal Setting depends on value of the outcome, task difficulty, specificity, and feedback \cite{LockeAPerformance}. In other words, a goal that has no appealing reward, is too easy or too hard, is vague, or has no feedback, is not an effective goal. In addition, fantasizing negatively about approaching the goal can increase stress and anxiety \cite{Elliot2007Approach-avoidance:Goals.,Elliot1997AvoidanceAnalysis}. For example, ``I will write to demonstrate my capability'' vs ``I will write to avoid being punished.'' Both might produce a similar outcome, but an avoidance goal might be more susceptible to procrastination behavior, because it is driven by negative thoughts. Selecting appropriate goals is a challenging task by itself. Thus, in this study we set our goal to be word-count, because it is specific, measurable, able to give real-time feedback, and nonjudgmental. We provide real-time feedback with a small progress bar showing the current number of words compared to the goal word count.  Users can set their own goal, but it is optional. In addition, we intentionally place the progress bar at a noticeable place at the top of the screen so that users can easily get access to the information (Figure \ref{fig-maiUI}. We hypothesize that this progress bar increases conscious motivation. 

\subsubsection{Writing Assistance System}
Procrastinators are poor at estimating the time necessary to finish a task \cite{ferrari_procrastination_2013}. Haycock suggests that splitting a task into small manageable chunks can help users get started \cite{Haycock1998ProcrastinationAnxiety}. With this system, we help users create a framework for their paper, as well as to divide a long paper into manageable sections. The application has a section panel to encourage users to create an outline. It allows users to only focus on one section at a time as opposed to scrolling through a long document. Users can search for certain sections via a search box, Figure \ref{fig-maiUI}. 


As discussed, a significant difficulty about writing is writer's block. Rose defines it as ``[...] an inability to begin or continue writing for reasons other than a lack of basic skill or commitment.'' He demonstrates that writer's block can be caused by lacking of creative ideas \cite{RoseWritersRhetoric}. Aren, in addition, defined writer's block as ``a condition producing a transitory inability to express one's thoughts on paper. It is characterized by feelings of frustration, rather than dread, hatred or panic'' \cite{Arem2011ConqueringAnxiety}. To help reduce writer's block, we used two text mining techniques: \textit{concept mining} and \textit{concept expansion}. Concept mining uses the initial content in a user's document to build a concept graph. These abstract concepts are used as keywords to search external sources and expand other related concepts in order to trigger creativity. In this paper, we used IBM Watson Concept Insight and IBM Watson Concept Extension API \cite{_concept_2016}. Once users had written 1000 words or more, we extracted the initial texts to find the top three concepts in the student's paper. Then, we used those concepts to find the top three TED talks that were most relevant to those topics. The system also uses the extracted concepts to search more adjacent concepts in the concept graph, providing additional TED talks about related areas. Finally, the system sends the result back to users via a clickable notification. We hypothesized that the system would help reduce anxiety and increase creativity, as well as provide an incentive to start writing early so that the concept map could be generated in time for the student to use the additional information. 

Finally, the writing assistance system creates custom question sets about the writing a student has generated. The goal was to have these questions stimulate creativity and create structure in the paper. These questions sets are ``wizard-of-ozed''---that is, the questions are generated by researchers, but students did not know if they came from a human or computer. The advantage of question sets over concept maps is that these question sets can be generated with less student writing and take less time for students to review. 

\section{Evaluation}
The custom text editor tracked word count, time spent typing, and documents versions for further analysis. To evaluate whether the design strategies affect users' procrastination behavior and satisfaction, we conducted a controlled trial experiment. The experiment consisted of two phases: a baseline phase and a follow-up phase. The baseline study was used to determine causes of procrastination and understand users' writing behavior without any notification system. In the follow-up experiment, notifications were added to the text editor. Moreover, subjects were grouped in the follow-up study by whether thay did or did not receive \textit{actioned notifications}. During both phases of the experiment we collected information about perceived procrastination behavior and usability of the custom text editor. We used the PASS survey and open-ended questions to determine the level of procrastination in our study population (discussed below). All participants were required to fill out the PASS survey and a system usability survey (System Usability Scale or SUS). All experiments were conducted with proper IRB approval.

\subsection{Baseline Study Design}
The objective of the baseline study was to identify potential procrastinators and non-procrastinators and preliminarily evaluate the editor for any major usability issues that could possibly contribute to procrastination. Based on data collected in the baseline study, we wanted to find groupings of participants for the follow-up experimental study (explained in the next section). It was also meant to help familiarize participants with using the editor.  All participants received the same version of the editor, but they did not receive any notifications. Participants were college students enrolled in a course on ubiquitous computing. They used the editor on a graded-3000-word-writing assignment about the history of UbiComp from Weiser's vision to present day. The writing also involved a creative component where students argued if certain elements from Weiser's vision had come to pass, been discarded, or evolved to different elements. The participants were a mix of undergraduate and graduate students. They used the editor for nine days leading up to the paper turn-in deadline.  

{\bf Grouping by Writing Performance}: The course instructor graded all paper assignments. Participants were separately divided into groups based on their assignment grade, \textbf{above average} and \textbf{below average}. It should be noted that all students showed mostly good writing skills and were motivated to perform well because the course was elective and presumably the students had interest in the subject matter. 

{\bf Grouping by Procrastination}: Once the baseline experiment had ended, two annotators independently reviewed figures of word count and writing time for each participant. Annotators did not have access to the performance grouping of the participants. The annotators reviewed the graphs and settled on two different criteria in order to divide the participants into procrastinators and non-procrastinators: the number of days before the deadline when a user started (Group X: More than 3 days before, Group Y: Day before) and the number of writing sessions and length of time spent on writing (Group A: Many sessions, Group B: A few long sessions or one long session). Using these criteria, the participants were grouped into high procrastination and low procrastination. The high procrastination group always started the day before (or day of) the deadline and spent 1 or 2 long sessions writing. Both researchers unanimously agreed on which students were in the high procrastination group. There was some disagreement on medium versus low procrastination for participants that started early, but only wrote a few long sessions; therefore, it was decided to group all participants who started early into the low procrastination group. 

{\bf Final Grouping for Follow-Up}: Finally, the researchers divided subjects using both the \textit{low/high} procrastination grouping and \textit{above/below} average groups as shown in Table \ref{fig-grouping_table}. For several participants, there was not enough writing data to divide them into high/low procrastination groups. In this case, they were placed in the ``no data'' group. This could occur, for example, for participants who never connected to the internet or whose firewall prevented the editor from sending word count and writing time information.

\begin{table}
\begin{center}
\begin{tabular}{ |c|c c| } 
 & Performance Level &\\
 \hline
 Procrastination level & Below Average & Above Average \\ 
 No Data & 2 & 3 \\ 
 Low & 3 & 5 \\ 
 High & 4 & 4 \\ 
 \hline
\end{tabular}
\end{center}
\caption{\label{fig-grouping_table}Breakdown of the number of participants by procrastination and performance level.}
\end{table}

It is interesting to note that, while most low procrastinators performed well on the assignment, many high procrastinators also did well. Furthermore, the distinguishing of ``procrastinator'' here is not a perfect measure because we cannot be sure whether or not any negative feelings or planned delays contributed to working on the paper the day before it was due. In other words, behavior that may have seemed like procrastination may have been planned by the student. The PASS survey results, however, support a conclusion that this student behavior was procrastination. 

The PASS (Procrastination Assessment Scale) \cite{mortazavi_psychometric_2015} survey instrument is well accepted in the psychology community for self-report of procrastination behavior. The PASS survey consists of two sections. The first section evaluates levels of procrastination. The section consists of eighteen items scaled 1-5. The second section identifies 13 reasons for procrastination. It consists of twenty-six items scaled 1-5. The first section of the PASS survey demonstrated procrastination levels (0-10). The high procrastination group had an average of 8.2 points (sd=0.45), and the low procrastination group had an average of 6.6 points (sd=1.52). Thus, the results of the PASS survey support the manual grouping that was based solely from writing behavior. 

To create two final experimental groups, we chose an equal number of students from each cell in Table \ref{fig-grouping_table}. This ensured that each experimental group had an approximately equal level of procrastination and writing performance. That is, each group was a representative sample of the class across procrastination level and writing performance. These experimental groups are designated as GRP1 and GRP2 for the remainder of this paper. 

\subsection{Baseline-Study Results}
\begin{figure}
\centering
\includegraphics[width=0.5\textwidth]{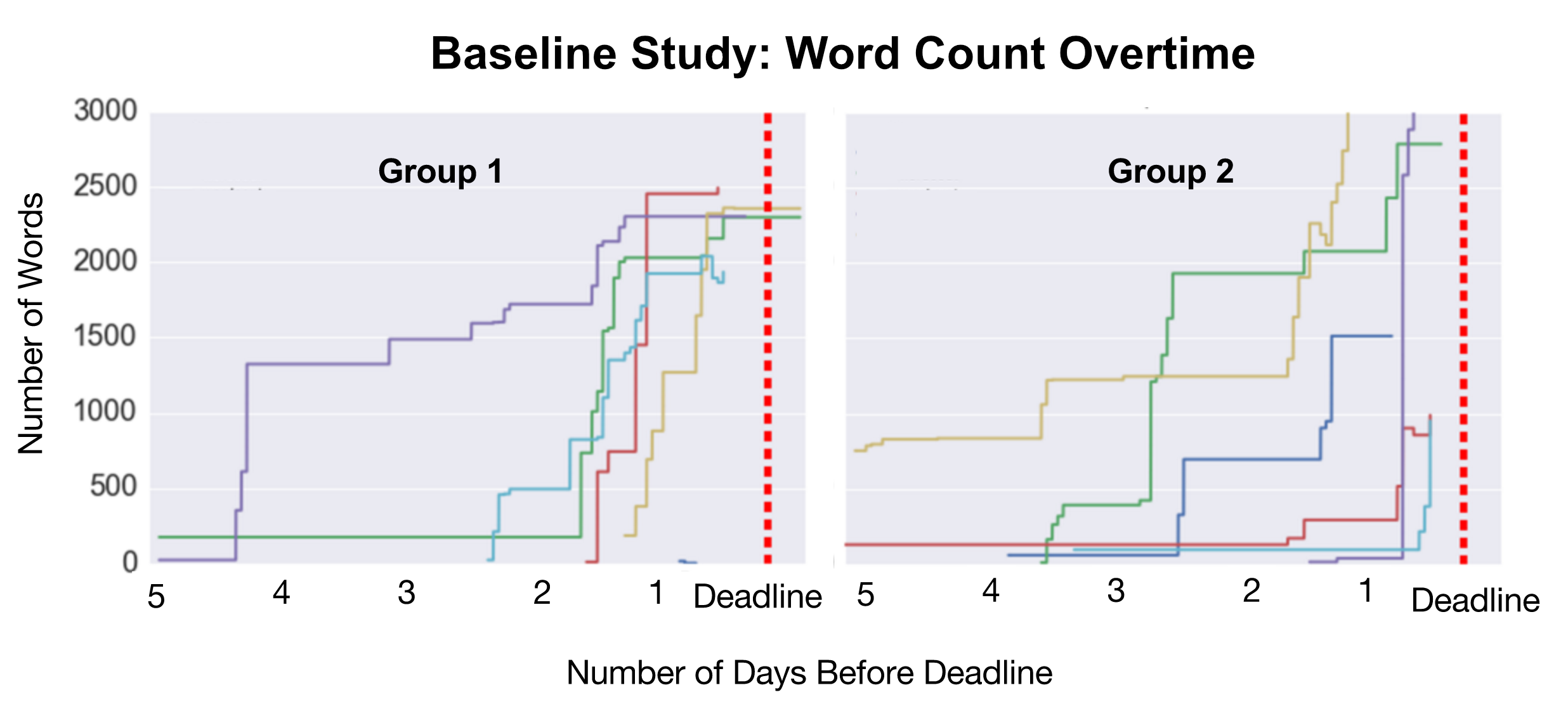}
\caption{\label{fig-wordcount_pilot}Number of words written graphed over time for baseline experiment. Each line represents a separate participant. }
\end{figure}

Although users were given nine days to work on the paper, Figure \ref{fig-wordcount_pilot} shows that on average, most participants started writing 2-3 days before the deadline. While GRP2 had two students that started on the assignment more than two days before the deadline (as opposed to GRP1 with only one student), both groups had a similar number of writing sessions per participant and a similar total writing time. GRP1 had an average of 2.8 writing sessions (sd=0.98), and GRP2 had an average of 2.0 writing sessions (sd=1.0). 

\begin{figure}
\centering
\includegraphics[width=0.5\textwidth]{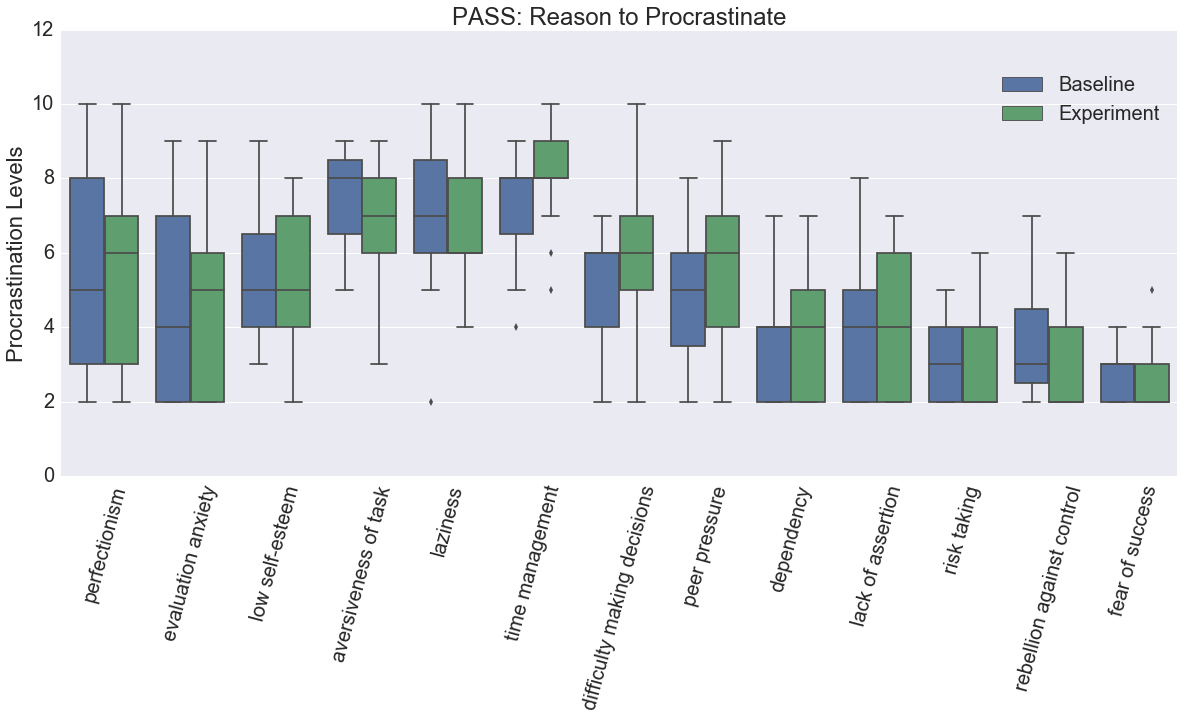}
\caption{\label{fig-passCmp}Comparing reasons to procrastinate between two groups for baseline and follow-up experiments.}
\end{figure}

Figure \ref{fig-passCmp} shows reasons of self-reported procrastination via the PASS survey for both baseline and follow-up studies. The top three reasons for procrastination were aversion to the task, laziness, and time management. Participants had varying reasons for aversion such as ``\textit{It's hard to put my thoughts onto paper}'' and ``\textit{I never know how [or] where to begin}.''

The result from SUS shows the software usability score is about average (mean=65.8, std=7.07). The reason most participants liked the editor is the cleanness and simplicity of the user interface, with quotes such as ``\textit{Very simple UI}'' and ``\textit{I liked the clean-ness of it}.''

The reasons for concern were software reliability and stability. As given by comments such as ``\textit{I just need to be guaranteed my work isn't going to be lost when the program crashes (which happened) otherwise I don't have enough confidence to use it}'' and ``\textit{crashing at the beginning freaking me out...}.''

Based on the given feedback, we implemented a set of upgrades to the editor before the second experiment. We retained the clean and simple UI while also minimizing user concerns by allowing users to export the document. We also conducted more extensive software testing to eliminate crashing. 

\section{Experimental Follow-Up Study Design}
To answer whether notifications help users get started in academic writing, the same 21 college students participated in a follow-up study. The students were assigned another graded 3000-word writing assignment. In this paper, students were asked to summarize two application areas of UbiComp and hypothesize about a research or class project that could contribute to one or both of these application areas. Students were given nine days to complete the assignment. As described, participants were divided into two representative groups that used the custom text editor, GRP1 and GRP2. The program recorded the following user actions related to writing: writing content, word count, typing time, received notifications time, and users' responses. For the second experiment, the word count and writing time sampling rates were increased from once per half hour to once per two minutes. This ensured a more reliable estimation of the length and number of writing sessions in which users engaged; moreover, the software is always running in the background, allowing us to continually collect data and push out notifications. Each group received a different set of features. GRP1 received standard clickable notifications only. In contrast, GRP2 received both standard clickable notifications and actioned notifications. We also sent out notifications that asked users their reasons for accepting or dismissing the notification. Finally, all participants again completed the Procrastination Assessment Scale for Students (PASS) after the experiment, and filled out the SUS survey after the experiment.

\section{Experimental Follow-Up Study Results}
Over the nine-day experiment period, we closely observed participants' behavior and provided a series of standard clickable and actioned notifications. Although all 21 participants agreed to take part in the study, 1 person was excluded because of low writing competency, 4 participants did not install the application, and 2 participants installed the software but never used it, in spite of sending several reminder emails. Therefore, 14 participants successfully installed and used the software. Half of the participants (n=7) were online regularly and the remaining 7 people were online a few hours per day. Fortunately, the number of disqualified participants was about the same in GRP1 and GRP2, leaving 7 people in each notification group.

In this study, we measured procrastination in two ways: writing statistics and standard self-reports. Writing statistics was measured by the number of sessions, starting writing date, and time spent writing per session. 

\begin{figure}
\centering
\includegraphics[width=0.5\textwidth]{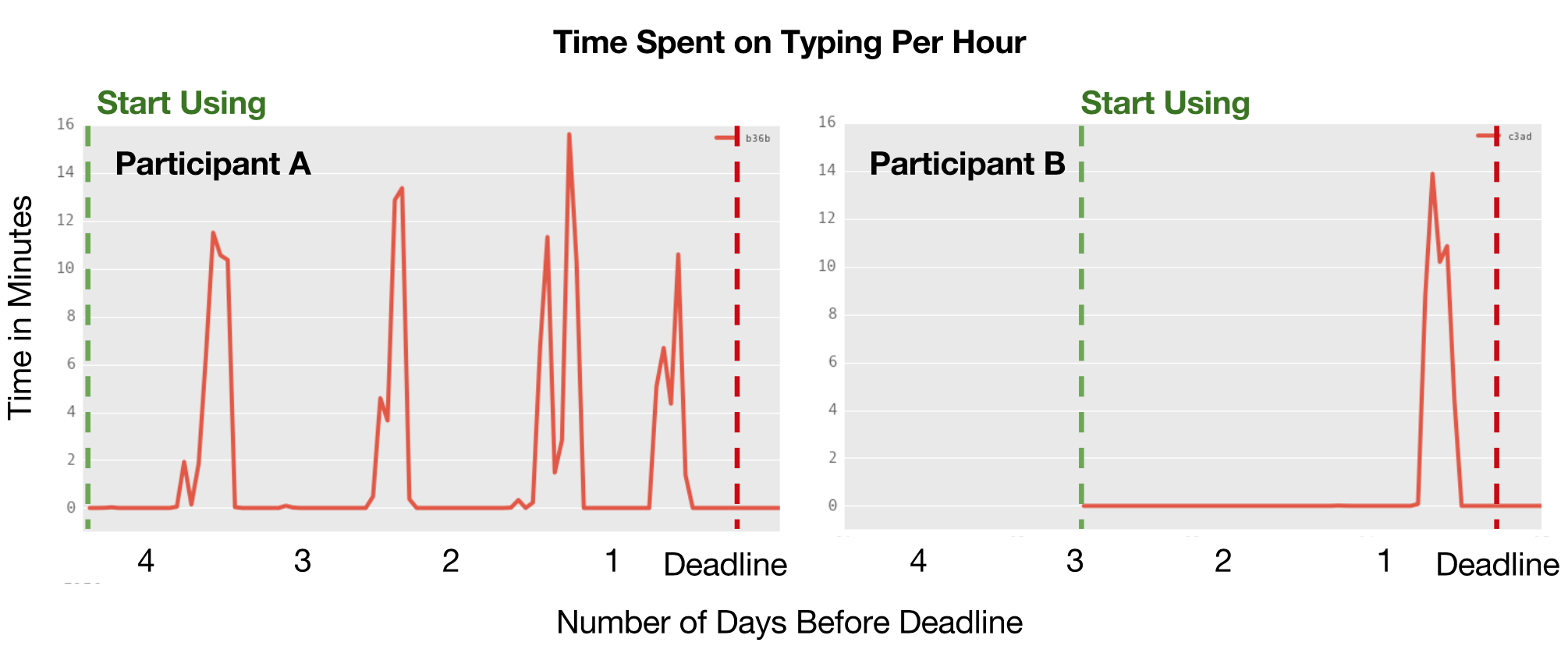}
\caption{\label{fig-timediff}Time spent writing continuously versus overall time before deadline for two example participants.}
\end{figure}

Figure \ref{fig-timediff} shows example of writing sessions for two users. The x-axis represents the number of days before the deadline, and the y-axis represents the amount of time spent writing continuously in minutes. These two participants represent two different behaviors from many participants: Some students had multiple short sessions, and some wrote in one long session close to the deadline. We used this data to identify the number of writing sessions for each participant. GRP1 had 2.29 sessions on average (sd=1.82, n=7), and GRP2 had 2.43 on average (sd=0.9, n=7). GRP1 wrote for 52.45 minutes on average (sd=44.11, n=7), and GRP2 wrote for 44.24 on average (sd=21.56, n=7). The data show that both groups spent short sessions at the beginning of the assignment and long sessions near the deadline.

To compare the performance of participants in both groups, the course instructor coded grades based upon two aspects: the novelty of the content (50 points), and how well they supported their arguments (50 points). To reduce the subjectivity of comparing point by point, we converted the range 0-100 to the discrete range F-A, and calculated the step difference. For example, if a participant got a B on the first paper and an A- on the second paper, we give him/her 2 steps difference (B,A,A-). GRP1 had a 0.75 step difference on average (sd=1.28, n=7) and GRP2 had a -0.29 step difference on average (sd=1.50, n=7).

\begin{figure}
\centering
\includegraphics[width=0.5\textwidth]{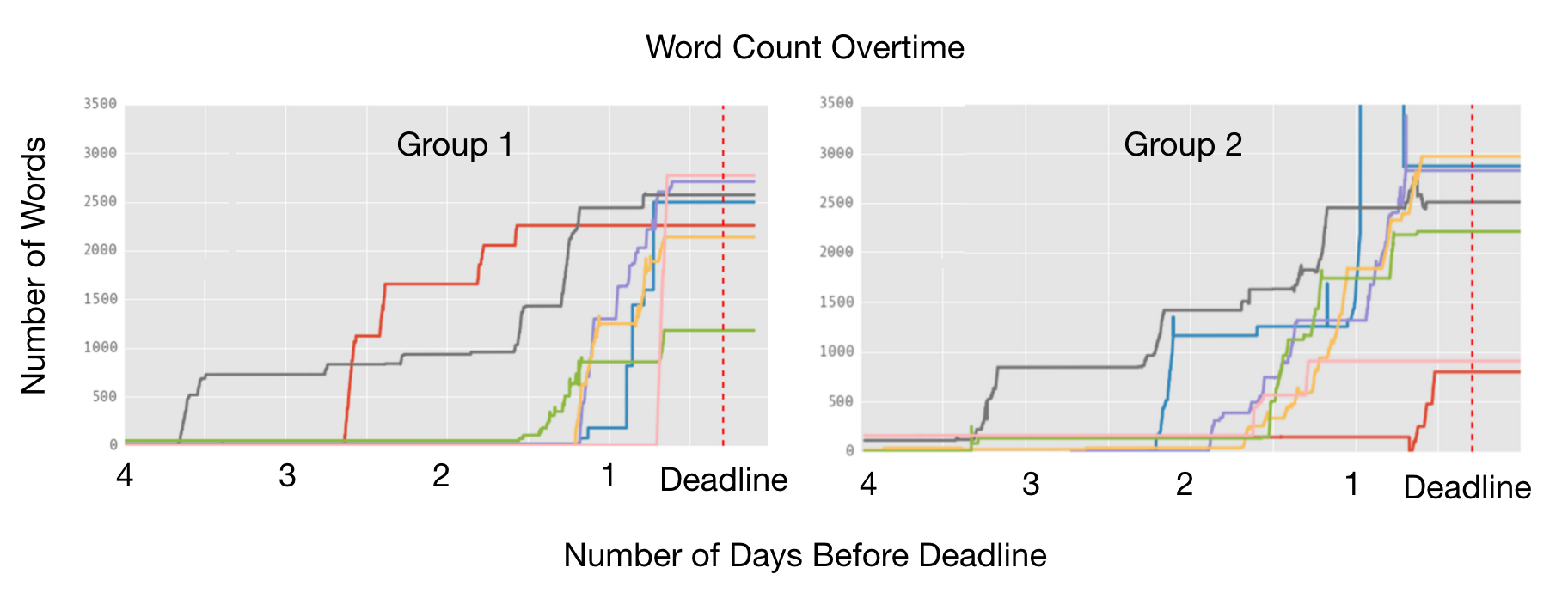}
\caption{\label{fig-wordcount_cmp}Word count versus time before deadline for each group in the follow-up experiment. Each line represents a separate participant.}
\end{figure}

Figure \ref{fig-wordcount_cmp} shows word count over the four days before the deadline. The x-axis indicates the number of days before the deadline and the y-axis shows the number of words over time. GRP2 does appear to have started slightly earlier than GRP1, but the small number of participants makes statistical testing inappropriate here. Qualitatively, GRP2 spent the full day and night before the deadline writing, whereas GRP1 mostly started writing the night before.

\begin{figure}
\centering
\includegraphics[width=0.5\textwidth]{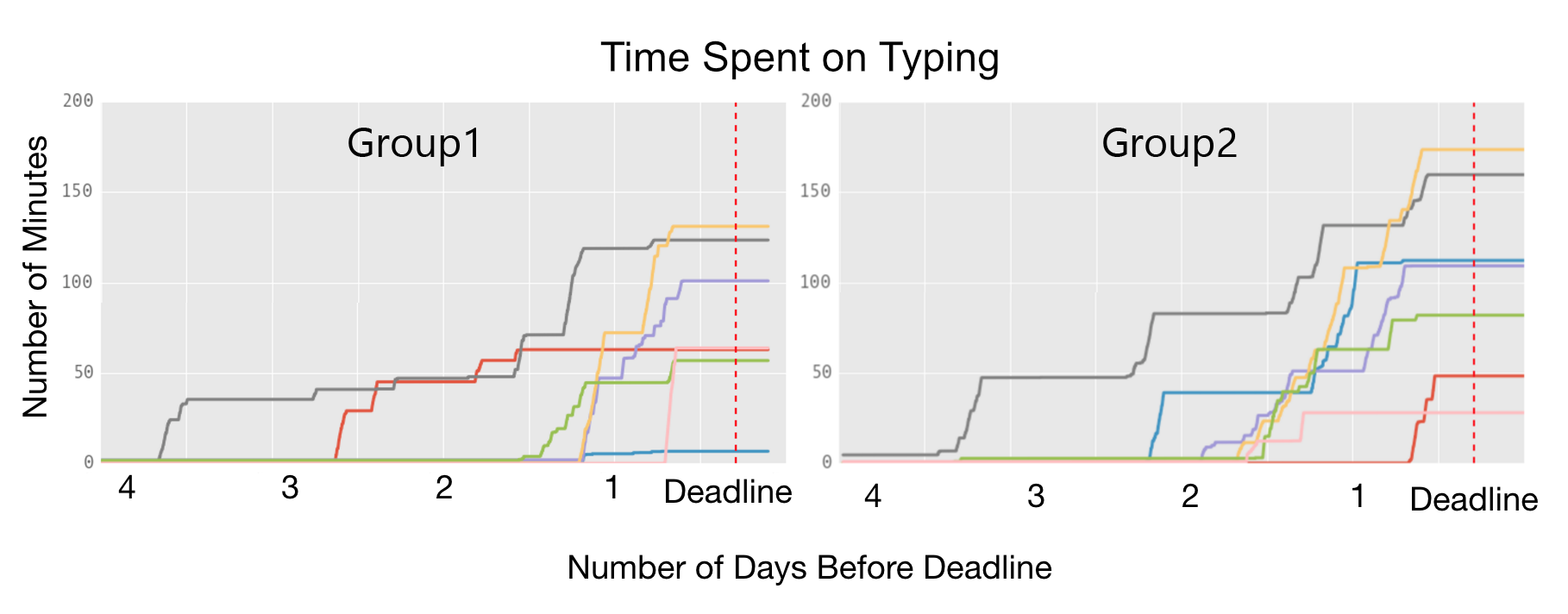}
\caption{\label{fig-timespend_cmp}Cumulative typing time versus time before deadline for each group in the follow-up experiment. Each line represents a separate participant. }
\end{figure}

Figure \ref{fig-timespend_cmp} shows the time spent on writing over 4 days before the deadline. The x-axis shows the number of days before the deadline and the y-axis shows the cumulative amount of time spent on typing in minutes. Despite clickable and actioned notifications being sent regularly for the full nine-day period, 5 people in GRP1 started around a day before the deadline, while 4 people in GRP2 started around 2 days before the deadline.

Participants received 143 notifications (118 clickable and 25 actioned notifications). All participants immediately dismissed all 118 clickable notifications. When we asked the reasons for dismissal, all participants claimed they were in the middle of something else (many students noted that they had mid-terms the week of the paper deadline). For the actioned notifications, on the other hand, 36\% were responded to (9 out of 25). Of the 9 notifications, 7 were related to outline-generation or content expansion. The last 2 notifications were short questions with a prompting text box. In these responses, the students only entered 2-3 words. When we asked the reason for such short answers, they also responded by saying that they were in the middle of another task. For students having trouble starting the assignment, we asked a random subset of the participants why they felt they did not get started quickly. The responses are low self-efficacy related, such as ``\textit{Not settle with the topic},'' ``\textit{not knowing what to write about},'' and ``\textit{Lack of ideas}.''

Our hypothesis was that actioned notifications would bypass negative reflection of the task by requiring users to make quick snippets of thought that helped to make writing more manageable; however, overusing notifications ended up increasing negative feelings. For example, ``\textit{It will be better if it knows when I am writing and then decide not to pop up questions, it is kind of a distraction}'' and ``\textit{Annoying notifications}.''

\begin{figure}
\centering
\includegraphics[width=0.5\textwidth]{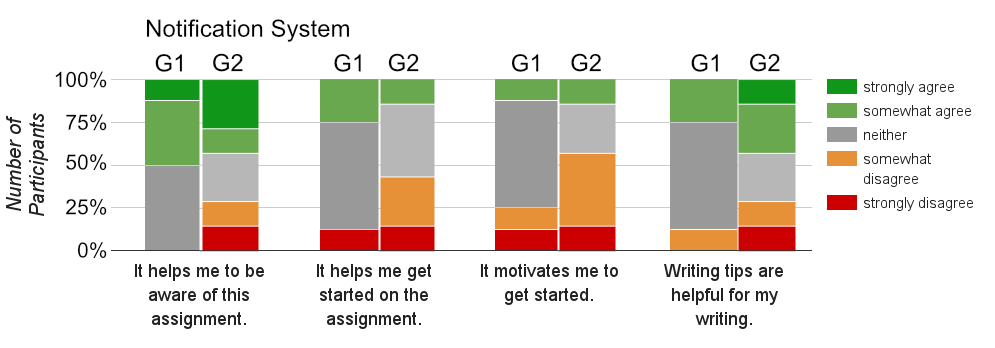}
\caption{\label{fig-notification_system}Survey response summary for the notification system.}
\end{figure}

We asked participants what features were most useful and influenced their decision making towards the writing tasks, summarized in Figure \ref{fig-notification_system}.
50\% of participants in GRP1 agreed that clickable notifications helped them be more aware of the task, 25\% said it helped them to get started, 12.5\% agreed that it motivated them, and 25\% thought writing tips were helpful. In contrast, in GRP2 (receiving clickable and actioned notifications) 43\% agreed that notifications helped them to be more aware of the task and thought writing tips were helpful. Only 14.29\% agreed that it helped them or motivated them to get started. The starkest difference between the groups, then, was their opinion of writing tips, where GRP1 was mostly neutral and GRP2 had stronger opinions about the tips, both positive and negative. 

\begin{figure}
\centering
\includegraphics[width=0.5\textwidth]{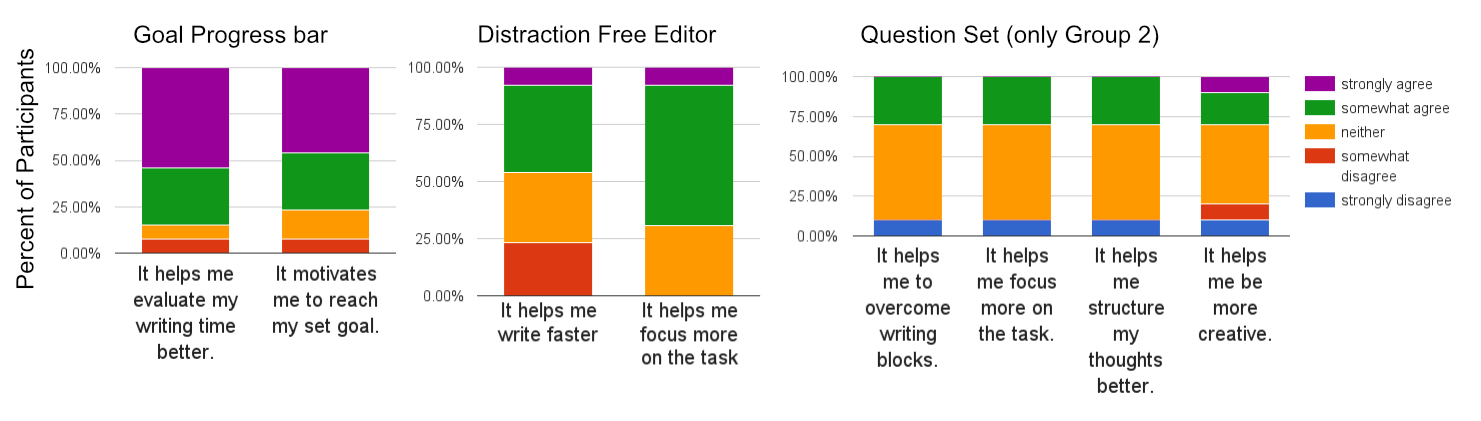}
\caption{\label{fig-distraction}Survey response summary for various elements of the custom text editor.}
\end{figure}

Both groups used distraction free mode. Figure \ref{fig-distraction} summarizes the perceptions of goal progress and distraction free mode. The result showed that 40\% agreed that it helped them write faster, and 60\% agreed that it helped them focus more on the task. Both groups used a goal progress bar indicator. The result showed that 80\% agreed that it helped them evaluate writing time better, and 73.33\% agreed that it motivated them to reach their set goal. Participants stated ``\textit{I like the word counting bar a lot!}'' and ``\textit{I like Word count and overall UI}.''

Only GRP2 received writing assistance notifications. The result showed that 30\% agreed it helped them overcome writer's block, 40\% agreed that it helped them focus more on the task, structure their thoughts better, and be more creative (Figure \ref{fig-distraction}). Participants had particularly strong opinions about these notifications: ``\textit{The automatic content generator learning system was absolutely amazing. I was shocked to see how accurate it was. It truly helped me when I was stuck and motivated me to keep going. If the negative aspects of the app are removed (listed below), I will absolutely use this app in the future},'' and ``\textit{I like the earlier planning questions to help me get started},'' and ``\textit{The planning questions, definitely helpful}.''

SUS scores for the updated the application had a mean of 60.7 points, with the standard deviation of 12.30 points (n=14). From qualitative comments, 11 people out of 14 participants liked the simplicity of user interface: ``\textit{I really enjoyed using it to write my paper. The UI was very simple and did not have a lot of distractions},'' ``\textit{Looks very clean},'' and ``\textit{it's very minimalistic and easy to use}.''

The dislike regarding the application stemmed from annoyance with the notifications and the fact that the application was always running. The memory usage of the application is about 60Mb (30Mb compressed), which is about 50\% less than Dropbox or Google Drive syncing service. The visibility that application was always running without users being able to control it, led to the increased negative feeling: ``\textit{I did not like having to keep the app running at all times}'' and ``\textit{It will better if it knows when I am writing and then decide not to pop up questions, it is kind of a distraction}.''

The outlining system also contributed to low usability scores. Many individuals did not understand how to use the section panel (despite attending a tutorial on using the application). A number of comments were similar to: ``[...] \textit{I was confused by the sections/files on the left side. What were they actually for?}''

\section{Discussion}
Our results show that there is very little to no difference regarding procrastination behavior after technological intervention; however, because we only captured the time and word count information, it was unclear when participants were conducting background research for their papers. Comparing the PASS self-reported surveys taken at the end of baseline and the end of the follow-up study, the data shows an insignificant difference between both baseline and follow-up procrastination level; however, Figure \ref{fig-passCmp} shows that the reason for procrastination shifted from ``aversion toward the task'' to ``time management''. The reason for this may be that the experimental study was the week before mid-term exams. Many course projects were due that week, including our writing assignment. Qualitative data also supports this hypothesis, with several participants mentioning mid-term exams. On the other hand, Writing assistants significantly decreased task aversiveness and difficulty in decision making compared to the baseline study (p<0.05), while other factors remained the same. This conclusion is also supported by qualitative data from participants. These results suggest that the perceived benefit of attending a notification must be considerable to be perceived as positive---the actioned notifications were the only notifications attended to, but required considerable time and user writing data to prepare.

\subsection{Notifications}
Although use of a notification is useful  for external triggers of procrastination, the notification not only triggers memory about the task,  but it might also trigger feelings associated with the task. If the task is aversive, those negative feelings (anxiety, guilt, or boredom) are more likely to be triggered. Thus, an aversive notification might be more susceptible to annoyance, guilt, and anger than the desirable ones. This is consistent with the basic psychological concept of conditioning, exemplified by a trigger (a door bell) associated with a particular outcome (food) \cite{seligman_prevention_1999}. In addition, our result implied that users may perceive an aversive notification as a reminder, regardless of motivational text written in the notification. We hypothesized that the tips and tricks notification may help users increase writing competency, leading to the desired amount of writing; however, it appears that users perceived the tips and tricks notifications as a disguised version of reminders. Thus, the number, frequency, and prominence of aversive notifications must be carefully considered. Some users may appreciate reminders; however, users should be given full control over whether or not they want to be notified, postpone the task, or not interact with it at all. Our experiment implies that only notifications by themselves are not enough to encourage users to get started on aversive tasks. 

\subsection{UI Elements and Writing Assistance}
Word count as a goal does indeed motivate users as we expected. Most users reported high satisfaction. The reasons that word count is so effective might be its specific, measurable, and nonjudgmental nature. More importantly, it gives users real-time feedback. They can see clearly  how much they have invested in their work. They focus on number of words, instead of focusing on perfecting those words. One may argue that the goal of academic writing is quality, not quantity. However, we argue that there will be no quality without some quantity. Quality has to start from some quantity and iterative improvement. If users fear starting, then they are less likely to produce any quality work \cite{Boice1990ProfessorsWriting}.

A number of participants liked the distraction free mode, and self-report shows that they feel more focused on the task when entering this mode. Since we did not track users outside our application, it remains unclear whether or not distraction free mode affected users on a behavioral level. 

The results imply that users appreciate systems that help them minimize time and effort required to finish a task. Writing assistant systems can reduce writers' anxiety. At the same time, concept extraction gives useful feedback to users. Concept extraction is a machine operation, so there is little to no fear of social judgment on the quality of the work. Adding semi-intelligent systems to help users finish aversive tasks easier and faster is a promising strategy to reduce procrastination and increase satisfaction. Human feedback may give strong social reinforcement. This power could increasing bursts of motivation more than any machine ever could. At the same time, social judgment can also paralyze users from getting started. Machine intelligence, in contrast, provides pure non-judgmental feedback. Users still get feedback while feeling safe from losing self-efficacy and social judgment. Balancing human and machine feedback could facilitate reducing procrastination behavior.

In our experimental design, we had no placebo group for this set of features; nonetheless, writing is a familiar behavior among college students. Most students have  experience with text editors without special features. Any self-evaluation is likely to be compared with this previous experience. Even so, this was not explicitly controlled for, so changes in attitude remain unclear.

\subsection{General Lessons Learned}
A key consideration for software designers should be data transparency. In this study, we were collecting basic user statistics such as word count, writing time, and notification reaction time. We told users exactly what data we collected including explicit written words through informed consent. In addition, writing a graded assignment is extremely important information for the user. Users have high expectations for software stability. We understood their concerns and told them that we have versioning systems. All data was backed up periodically; nonetheless, some users still showed some level of discomfort with using it. This might be caused by inability to manage their own data. Thus, offering data transparency appropriately may create more trust among users and researchers.

\subsection{Summary of Lessons Learned}
\begin{enumerate}
\item If the goal is to create an immediate action, all blockers have to be identified and eliminated before sending triggers.
\item Notifications also trigger emotion that are associated with the task. Be mindful before using them.
\item Notifications are effective to decrease user's burden of keeping track of a list of tasks, but they are less useful in persuading actions that are aversive.
\item Users may perceive aversive notifications as a negative reminder, regardless of  motivational text written in the notification.
\item Showing users proper, measurable, and specific goal helps users evaluate their work better.
\item It is a good idea to keep users' autonomy in mind. If you decide to use notifications, make sure you provide options for users to opt-out.
\item Adding semi-intelligent systems to help users finish aversive tasks easier and faster is a promising strategy to reduce procrastination and increase satisfaction. 
\item Questions are useful for users who are already motivated to answer them; however, it backfires for users who do not. Before asking questions, it is a good idea to make sure users have enough answers or motivation to answer them.
\item Users are concerned for their own data. Considering data transparency helps users trust the service and increases satisfaction overall.
\end{enumerate}

\section{Conclusion}
We discussed the current psychological literature regarding procrastination and evaluated various technological interventions to decrease writing procrastination among college students. We also outlined the challenges and lessons learned through conducting procrastination research. Notifications used in this papers did little to decrease procrastination behavior; moreover, users who get more notifications have lower satisfaction than other peers. Helping users clear their motivation blockers is the first step in performing any task. Goal Setting Theory has proven effective in increasing motivation. Using machine learning aids can decrease the aversion toward the tasks. Thus, providing tools for making aversive tasks easier and less fearful is a promising strategy to decrease procrastination, but must be carefully applied---especially when employing notifications as motivators.

%
%
%
%
%
\balance{}

\bibliographystyle{SIGCHI-Reference-Format}
\bibliography{mai,Mendeley}

\end{document}